\documentclass[pdftex,twocolumn,showpacs,preprintnumbers,floatfix,prl]{revtex4}

\usepackage{amsmath,amssymb}
\usepackage{bm}
\usepackage{dcolumn}
\usepackage{graphicx}
\usepackage{hyperref}
\usepackage{rotating}
\usepackage{times}

\newcommand{\fifD}{$^{15}$ND$_3$}
\newcommand{\forD}{$^{14}$ND$_3$}

\newcommand{\forH}{$^{14}$NH$_3$}
\newcommand{\JK}{$\left|J,K\right>=\left|1,1\right>$}

\begin{document}
\title{Nonadiabatic transitions in electrostatically trapped ammonia molecules}
\author{Moritz Kirste}
\author{Boris G. Sartakov$^1$}
\author{Melanie Schnell$^*$}
\author{Gerard Meijer}
\affiliation{Fritz-Haber-Institut der Max-Planck-Gesellschaft, Faradayweg 4-6, D-14195 Berlin, Germany \\
$^1$General Physics Institute RAS, Vavilov Street 38, 119991 Moscow, Russia}
\date{\today}

\begin{abstract}
Nonadiabatic transitions are known to be major loss channels for
atoms in magnetic traps, but have thus far not been experimentally reported
upon for trapped molecules. We have observed and quantified 
losses due to nonadiabatic transitions for three isotopologues
of ammonia in electrostatic traps, by comparing the trapping times in
traps with a zero and a non-zero electric field at the center.
Nonadiabatic transitions are seen to dominate the overall loss rate even 
for the present samples that are at relatively high temperatures of 30
mK. It is anticipated that losses
due to nonadiabatic transitions in electric fields are omnipresent in ongoing 
experiments on cold molecules. 
\end{abstract}

\pacs{37.10.Pq, 37.10.Mn, 31.50.Gh}
\maketitle

The recent development of a large variety of methods and devices for the
manipulation and trapping of neutral polar molecules offers new opportunities 
for molecular physics experiments \cite{Meerakker:NatPhys4:595}. 
Decelerated beams and trapped samples of polar molecules can be used to 
study intrinsic molecular properties, such as energy level splittings 
\cite{Veldhoven:EPJD31:337,Hudson:PRL96:143004} and 
lifetimes of metastable states \cite{Meerakker:PRL95:013003,Campbell:PRL100:083003}, 
with unprecedented precision. Once the densities of the trapped molecules 
become high enough and their temperatures become low enough, the 
inter-molecular interactions are anticipated to enable interesting new 
studies and applications \cite{Krems:IRPC24:99,Andre:Natphys2:636}. 
For all these studies it is not only of importance to increase the phase-space
density of the trapped molecules but also to increase the time during which 
the molecules can stay confined in the trap, i.e., to reduce the trap loss processes.
Neutral polar molecules in low-field seeking states 
are routinely trapped in magnetostatic or electrostatic traps by exploiting the 
Zeeman or Stark effect, respectively \cite{Weinstein:Nat395:148,Bethlem:Nature406:491}. 
These traps typically exhibit zero field at the trap center. Within a 
certain area around the trap center, the trapped molecules can undergo 
nonadiabatic transitions, widely also referred to as spin flip or Majorana transitions, 
from a trapped state into a state in which the molecules are no longer trapped. 
In atomic physics, nonadiabatic transitions in quadrupole magnetic traps seriously 
hindered the generation of the first Bose-Einstein condensates, as these spin flips
made it impossible to reach the required ultracold regime \cite{Ketterle:RMP74:1131,Cornell:RMP74:875}. 
The trap losses associated with the presence of the zero field at the trap center
were eliminated by implementing the TOP (time-averaged, orbiting potential) trap
on the one hand \cite{Petrich:PRL74:3352} and by keeping the
atoms away from the trap center with an optical plug on the other
hand \cite{Davis:PRL75:3969}. Generally, trap loss due to nonadiabatic 
transitions can be completely suppressed by creating a non-zero field 
minimum in the trap center. Already in 1962, Ioffe introduced a special variation 
of a magnetostatic trap with a field offset in the center for nuclear physics 
experiments \cite{Gott:NuclearFusion3:1045}. Pritchard suggested in 1983 to 
use such a trap for the confinement of neutral atoms \cite{Pritchard:PRL51:1336}. 
This type of magnetostatic trap is now widely known as the Ioffe-Pritchard (IP) trap.

For trapped polar molecules, losses due to nonadiabatic transitions have
not been experimentally reported upon yet. Nevertheless,
several possible geometries for an electrostatic analogue of an IP-type trap 
have been suggested to prevent these possible losses, such as a chain-linked 
trap \cite{Shafer-Ray:PRA67:045401} or a six-wire trap \cite{Xu:thesis:2001}.
More recently, a trap with an electric field offset in the center has been 
demonstrated for Rydberg atoms \cite{Hogan:PRL100:043001}. Theoretical 
studies on the loss of molecules from purely electrostatic traps \cite{Kajita:EPJD38:315} 
or, more generally, from magneto-electrostatic traps \cite{Lara:PRA78:033433} 
due to nonadiabatic transitions have also recently appeared. In both studies
it was concluded that -- although this can be different for any particular 
molecule -- the loss rate is negligible at the milli-Kelvin temperatures 
that are currently achieved in electrostatic traps.

In this Letter, we experimentally study trap losses due to nonadiabatic 
transitions in electrostatically trapped ammonia molecules. We quantify 
this trap loss mechanism from measurements of the trapping times
in an IP-type electrostatic trap with either a zero field or an offset field at the center, 
under otherwise identical conditions. Moreover, we measured the trapping times 
for three isotopologues of ammonia, namely \forH, \forD~and \fifD. These molecules 
have subtle differences in their energy level structure, resulting in different 
probabilities for nonadiabatic transitions. Contrary to the theoretical expectations, 
the results demonstrate the enormous importance of trap losses due to nonadiabatic 
transitions even at the present 30 mK temperatures. 

\begin{figure}
   \begin{center}
   \includegraphics[width=\linewidth]{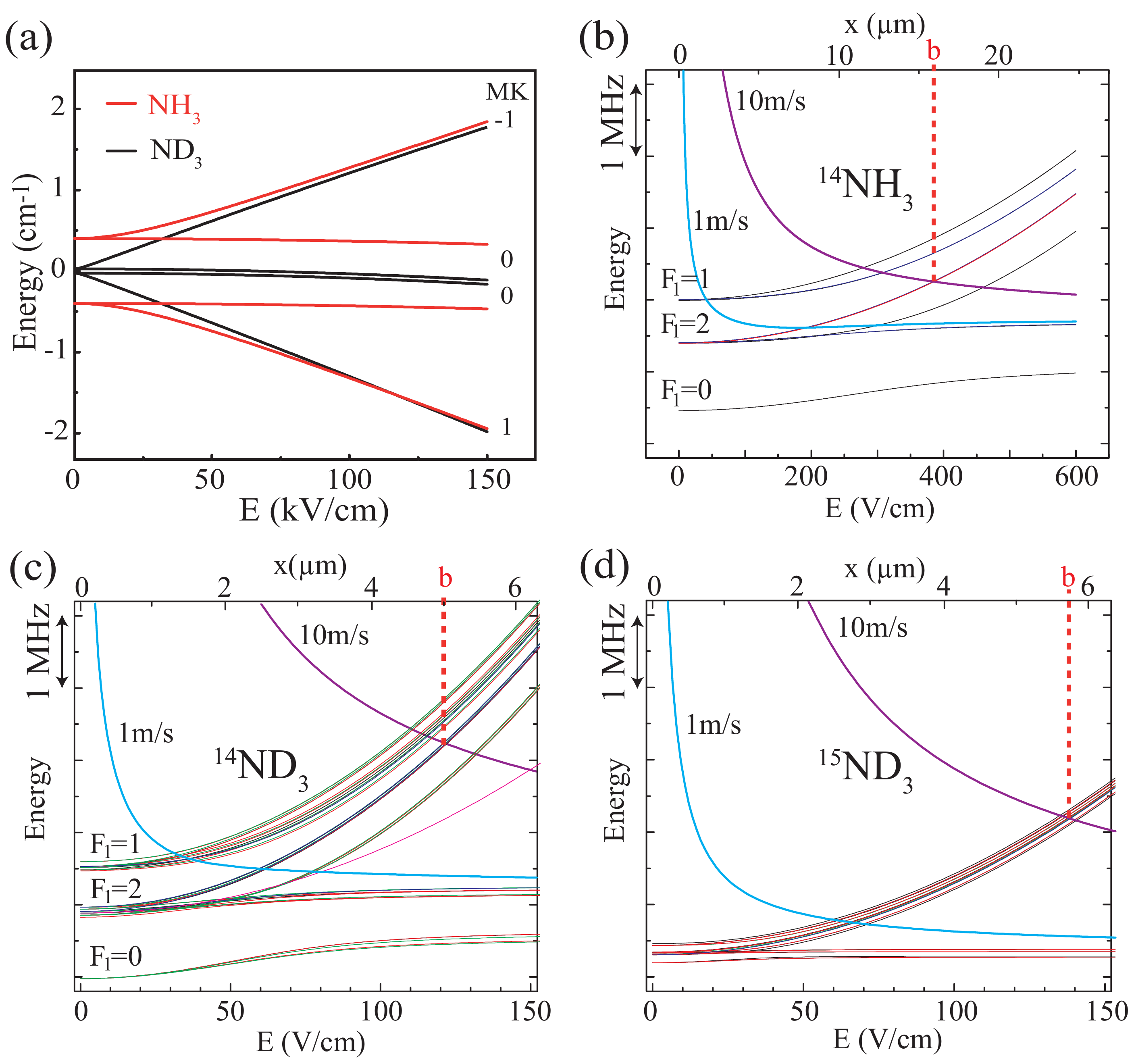} 
   \caption{(Color online) Energy of the \JK~level of \forH~ (dashed
      line) and \forD~ (straight line) in fields up to
   150 kV/cm (a). In panel (b), (c) and (d) the Stark shift of the upper inversion
   doublet hyperfine components in low electric fields is shown on expanded scales for 
   \forH, \forD~ and \fifD, respectively. For the calculation of the
   electric field dependence of the hyperfine components, see Ref. \cite{Veldhoven:PRA66:032501}. 
   The $v/x$ curves indicate the
   rate of change of the direction of the electric field for a molecule with
   velocity $v$ (1 and 10~m/s) passing the trap center at a distance $x$ (upper
   axis).}
   \label{fig:Stark}
   \end{center}
\end{figure}

Figure \ref{fig:Stark}(a) shows the energy of the \JK~level of NH$_3$ and 
ND$_3$ in electric fields up to 150~kV/cm. The upper (lower) inversion doublet 
components of this level are seen to split into a low-field (high-field) 
seeking set of hyperfine levels labeled $MK$=-1 ($MK$=+1) and into a set of 
hyperfine levels labeled $MK$=0 that are basically not shifted in the fields. 
The zero-field inversion splitting of \forH~is with W$_{inv}$=23.7 GHz 
considerably larger than the W$_{inv}$=1.59~GHz of \forD. At low electric 
fields, the Stark shift of the $MK$=-1 set of levels is 
proportional to $E^2$/W$_{inv}$, and is therefore considerably less for 
\forH~than for \forD~in a given electric field with magnitude $E$. 
In Figure \ref{fig:Stark}(b) the Stark shift of the upper inversion
doublet component of \forH~is shown for low electric fields on an expanded 
scale, such that the individual hyperfine levels can be recognized. The 
quadrupole interaction of the $^{14}$N nucleus causes the observed 
main splitting in three groups of levels, labeled with the quantum
number $F_1$ \cite{Kukolich:PR156:83}.
The nuclear spins of the hydrogen atoms cause small additional splittings. 
In the electric field, the overall splitting behavior into (almost) unshifted 
levels belonging to $MK$=0 and levels that shift quadratically with $E$, 
belonging to $MK$=-1, is recognized. The individual 
hyperfine levels have a definite angular momentum projection quantum number
$M_F$, where $\vec{F}$ is the total angular momentum. As the electric 
field scales linearly with the distance from the center in a quadrupole electrostatic trap, a
distance scale ($x$, in $\mu$m) appropriate for the trap used in our experiment 
is indicated on top of the Figure as well. In the lower two panels of Figure \ref{fig:Stark} 
the behavior of the hyperfine levels of \forD~and \fifD~in electric fields 
is shown \cite{Veldhoven:PRA66:032501}. Note that the electric field scale, 
and thereby the distance scale, is now a factor four smaller than 
for \forH. For \forD~ and \fifD, the magnitude of the inversion splitting is
quite similar (1.59~GHz and 1.43~GHz, respectively) but their hyperfine 
structure is rather different. As the deuterium atoms have a nuclear spin
of one, the hyperfine structure in the deuterated isotopologues of ammonia
is generally quite complicated. In \forD, the hyperfine levels again split in three
groups of levels due to the quadrupole coupling of the $^{14}$N nucleus 
\cite{Veldhoven:PRA66:032501}. The $^{15}$N nucleus has no quadrupole 
moment, resulting in less (11, instead of 16 for \forD) hyperfine levels, but these are 
closer spaced and without a clear sub-structure for \fifD~\cite{Veldhoven:EPJD31:337}.

Polar molecules are oriented by the local electric fields inside the trap, and
they normally remain in the same quantum state and adiabatically follow the 
field while moving through the trap. The adiabatic eigenstates can be 
quantized with respect to the axis along the electric field vector $\vec{E}$ and 
can thus be assigned the quantum number $M_F$. Nonadiabatic transitions 
can occur when the molecules can not follow the rapid change of the direction 
of the electric field when they pass with velocity $v$ at a close distance $x$ 
to the region of zero electric field at the trap center. When this rate of change, 
given by $v/x$, is larger than the energy difference between levels in the 
molecule, nonadiabatic transitions can occur. As $\vec{v}$
is in general not oriented along $\vec{E}$, nonadiabatic
transitions with $\Delta M_F$=0 as well as with $\Delta M_F=\pm$1 will
be possible. In the case of ammonia, nonadiabatic transitions from the 
$MK$=-1 to the $MK$=0 set of hyperfine levels will cause trap loss. The 
frequency of these transitions at a distance $x$ from the trap center is 
proportional to $x^2/W_{inv}$. From this
it is seen that nonadiabatic transitions can occur when the molecules
come closer to the trap center than a typical distance $b$ $\propto$ 
$(W_{inv}v)^{1/3}$. In Figure \ref{fig:Stark}(b)-(d) curves of $v/x$ are 
indicated (in MHz) for two different velocities ($v$=1, 10 m/s), relative to the 
highest frequency $MK$=0 hyperfine level. The crossing point of these curves
with each of the $MK$=-1 hyperfine levels gives the corresponding distance
$b$ for nonadiabatic losses from that particular hyperfine level. Obviously,
when there is a large non-zero field at the center of the trap, nonadiabatic
transitions from the $MK$=-1 to the $MK$=0 levels will no longer be possible. 
Nonadiabatic transitions within the $MK$=-1 set of hyperfine levels can
still occur to some extent, although also these will be largely suppressed
as only the magnitude of the electric field, but no longer the direction, changes
rapidly when the molecules pass through the center of the trap.

In the experiment ({\it vide infra}), we detect molecules in all low-field seeking 
$MK$=-1 hyperfine levels simultanuously. It is evident from inspection of the Stark 
curves shown in Figure \ref{fig:Stark} that each of these levels will have a 
different nonadiabatic transition rate. Therefore, when the trap loss is dominated 
by nonadiabatic transitions, a multi-exponential decay is expected to be observed.
The rate for nonadiabatic transitions depends on how often the molecule passes
through the volume with cross-section $b^2$ near the center of the trap. This
rate can be approximated by $f_{osc}(b/b_{0})^2$, where $f_{osc}$ is the oscillation
frequency of the molecules in the trap and $b_0$ is the mean value of
the impact parameter of the trajectories of the molecules.
With a typical value for $f_{osc}$ of 1-2~kHz and with $b_0$ on the order of 0.2~mm, 
a value of $b$ around 5~$\mu$m will lead to a loss rate due to nonadiabatic 
transitions on the order of 1~Hz. 

\begin{figure}
   \begin{center}
   \includegraphics[width=\linewidth]{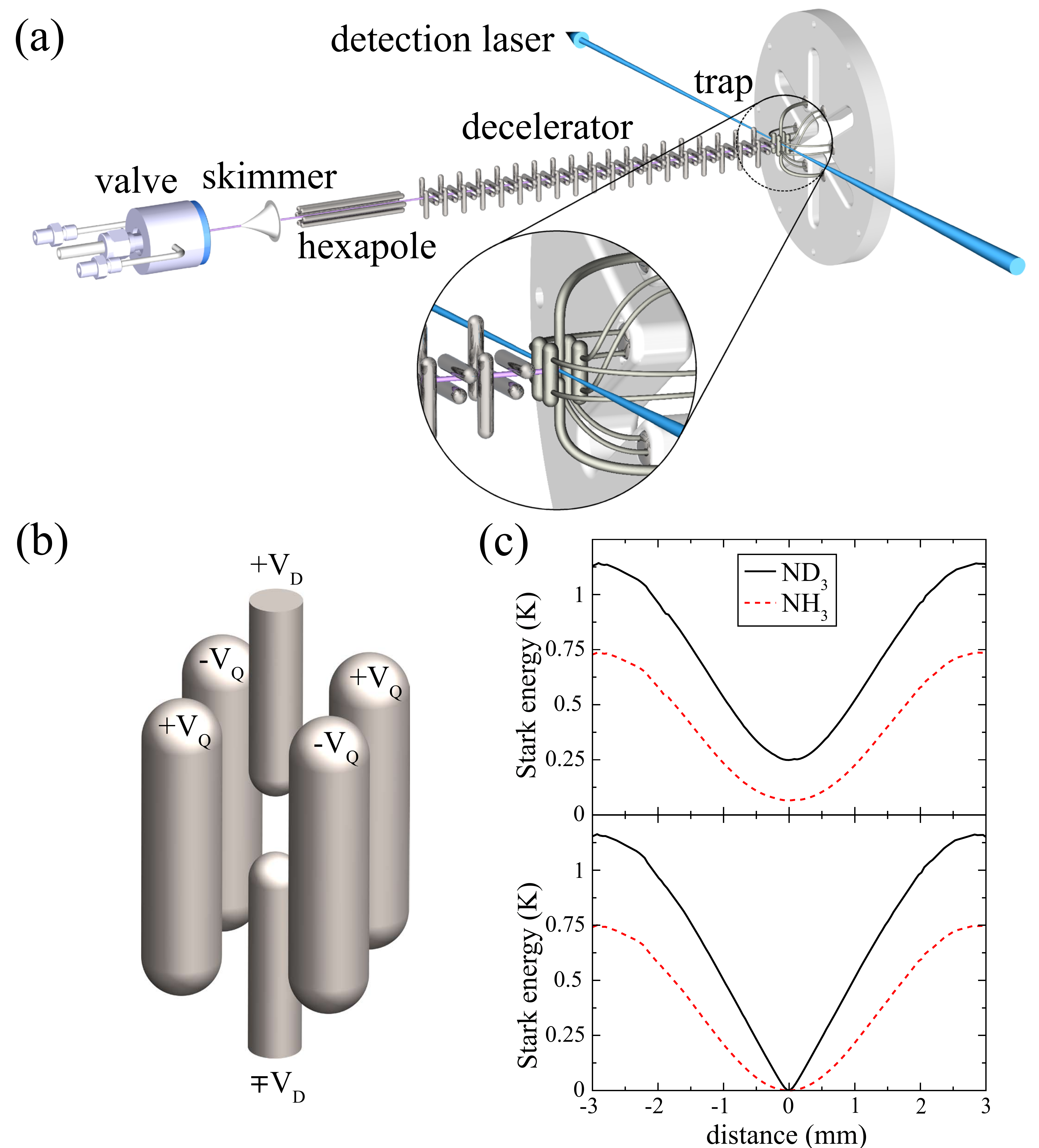} 
   \caption{Scheme of the experimental setup, with an expanded view 
   of the IP-type electrostatic trap. The voltages applied 
   to the six cylindrical electrodes are shown (b), together with the resulting
   potential energies for NH$_3$ (dashed line) and ND$_3$ (straight line) along a line in the
   horizontal symmetry plane for a non-zero (top) and a
   zero (bottom) electric field at the trap center (c).}
   \label{fig:exp}
   \end{center}
\end{figure}

The experimental setup, together with an expanded view of 
the IP-type electrostatic trap, is schematically shown in Figure~\ref{fig:exp}(a). 
A detailed description of the molecular beam machine, and in particular of the 
deceleration of a beam of ammonia molecules, is given elsewhere \cite{Bethlem:PRA65:053416}.
Decelerated packets of either \forH, \forD, or \fifD~molecules leave the decelerator 
in the low-field seeking $MK$=-1 hyperfine levels of the \JK~level with a mean velocity of around 
20 m/s, and with a full width at half maximum (FWHM) velocity spread of about
10~m/s. The decelerated packet contains
appproximately 10$^5$-10$^6$ molecules and has a spatial extent of about
2 mm along the molecular beam axis and is about 2x2 mm$^2$ in the transverse 
direction at the exit of the decelerator \cite{Bethlem:PRA65:053416}. The molecules are decelerated to a mean
velocity around zero m/s upon entering the trap, which is centered 8~mm behind the 
decelerator. After a certain trapping time, the trap is turned off and the molecules 
are state-selectively detected using a laser based ionization detection scheme.

The trap consists of six cylindrical electrodes, shown more clearly in 
Figure \ref{fig:exp}(b). To the four outer electrodes (3~mm diameter) positive 
and negative voltages ($\pm V_Q$) are applied such as to generate a quadrupole field. 
The two electrodes (2~mm diameter) centered on the symmetry axis of the trap
enable to create an additional dipole field. When voltages of opposite polarity
($\pm V_D$) are applied to these electrodes, a non-zero electric field is generated
at the center of the trap; by applying the same polarity ($+V_D$) to the center
electrodes, a standard quadrupole trap with a zero field at the center is obtained. 
In Figure~\ref{fig:exp}(c) the potential energy for the $MK$=-1 levels
of NH$_3$ and ND$_3$ in the trap is shown along
a line in the horizontal symmetry plane, both for a zero
(bottom) and a non-zero field (top) at the trap center. In the latter case, the trap 
depth is actually limited by the occurence of four saddle points of the electric fields in 
between the electrodes. For V$_Q$=10~kV and V$_D$=3~kV,
the offset electric field at the center is 16 kV/cm and the trap depth is about 290 mK 
for ND$_3$ and 130 mK for NH$_3$. With no electric field at the center, the trap
is considerably deeper, about 1.2 K for ND$_3$ and 750 mK for NH$_3$. 

\begin{figure}
   \begin{center}
   \includegraphics[width=0.8\linewidth]{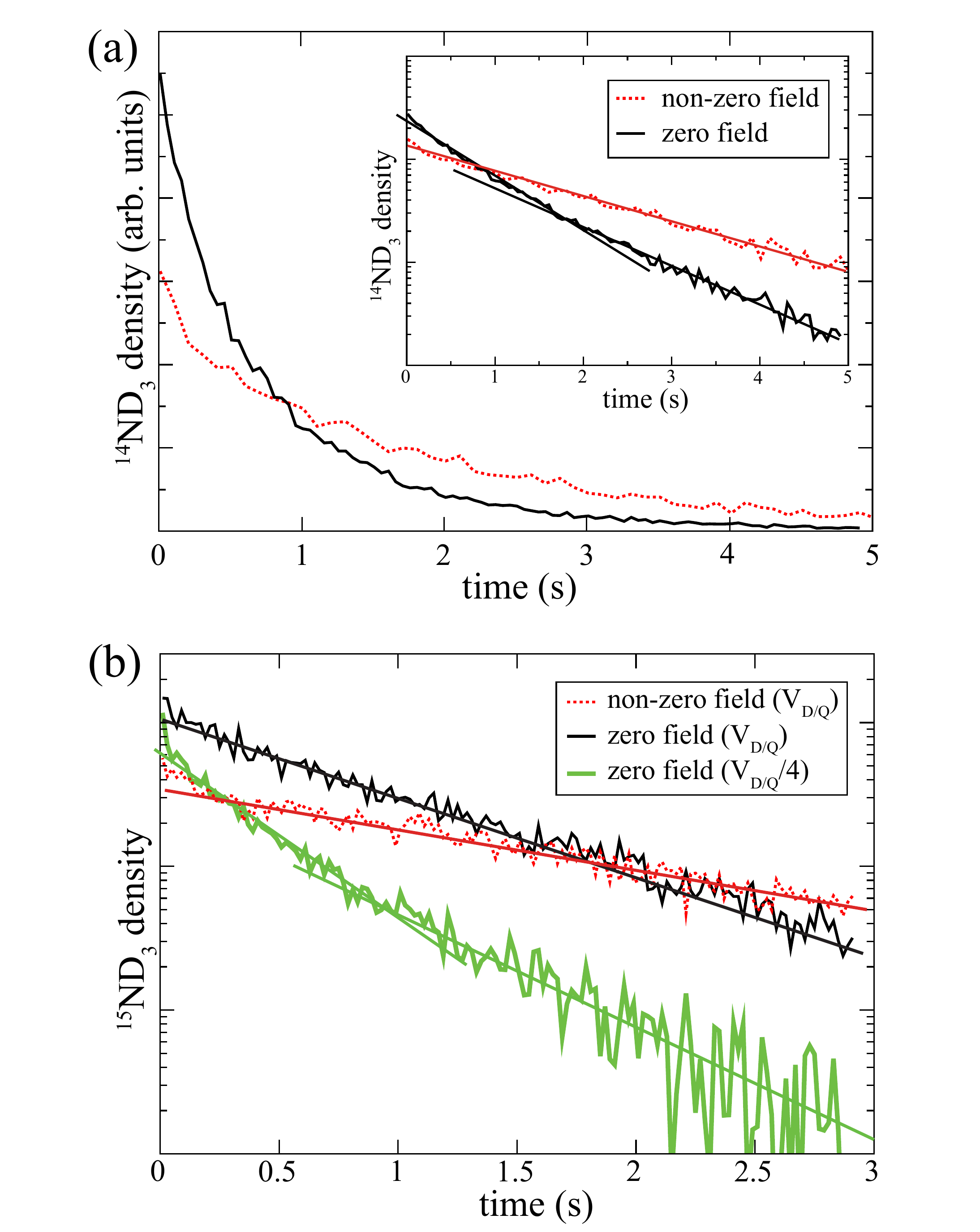} 
   \caption{(Color online) (a) Lifetime measurements for \forD~in the two trap
      configurations, both on a linear and  on a logarithmic scale (inset). 
      In the inset, straight lines are included to guide the eye. 
      (b) Lifetime measurements for \fifD~with original (V$_{D/Q}$) and
      reduced (V$_{D/Q}$/4) trap voltages on a logarithmic scale. The
      measurements have been performed with intervals of 20~ms and with 16
      averages per timepoint. The non-zero field in the trap center is
      16~kV/cm for all measurements shown.}
   \label{fig:14}
   \end{center}
\end{figure}

Figure~\ref{fig:14}(a) shows measurements of the density of \forD~molecules in
the electrostatic trap as a function of the time during which the trap has been 
switched on, both for zero and non-zero field at the center. Although the number
density in the zero field trap is originally higher due to a better trap loading, it is 
also seen to decay significantly faster. In both cases, the trap is considerably 
deeper than the temperature of the trapped ensemble of molecules; the latter
has been determined from expansion measurements to be around 30 mK.
The molecules can therefore only leave the trap via collisions, 
via optical pumping due to blackbody radiation or via
nonadiabatic transitions. Losses due to collisions can either originate from 
(in)elastic collisions of the trapped molecules with background gas or from cold
collisions amongst the trapped molecules. At the present densities in the trap 
(10$^7$-10$^8$ cm$^{-3}$), collisions of the trapped molecules with each other 
can be neglected. We have experimentally verified this by changing the
number density of trapped molecules over one order of magnitude, without observing
any change in the trapping time. The \forD~molecules in the low-field seeking 
$MK$=-1 hyperfine levels can be optically pumped to excited ro-vibrational levels 
due to blackbody radiation from the room-temperature environment, leading to a 
calculated finite lifetime of about 7 s \cite{Hoekstra:PRL98:133001}. The 
purely single exponential decay of the number of \forD~molecules in the trap with the 
non-zero field has a 1/e time constant of 1.9 s. This decay time results from the 
combination of optical pumping due to blackbody radiation and (in)elastic collisions 
with background gas (background pressure 10$^{-8}$ mbar). By simply changing 
the polarity on one of the center 
electrodes of the trap, under otherwise identical conditions, a much faster and 
multi-exponential decay is observed (Figure \ref{fig:14}(a)); the additional trap loss observed in the 
trap with zero electric field at the center is solely due to nonadiabatic transitions.
From these measurements, it appears that the nonadiabatic transitions for \forD~can
be approximated with only two time constants. Referring to Figure \ref{fig:Stark}(c), the
fast decay can be attributed to nonadiabatic transitions within the $F_1$=2 set of
hyperfine levels, whereas the slow decay can be attributed to nonadiabatic transitions
from the $F_1$=1 to the $F_1$=2 set of hyperfine levels. When the nonadiabatic 
transitions are modelled in this way (with the pre-factors for the two exponentials 
determined by the number of $M_F$ levels), time constants of about 0.25 s and 2.0 s, 
respectively, are extracted for these processes. The errors on these 1/e
time constants are less than 20~ms.

We have performed a similar series of experiments for \forH~(data not shown); this
is actually the first time that this ammonia isotopologue has been trapped at all.
In the zero field trap, a bi-exponential decay is also observed for \forH, which
can be explained in the same way as for \forD. After correction for the overall 
decay as measured for \forH~in the trap with the offset field, the time constants 
for the fast and the slow process are determined as 0.13 s and 3.5 s. The faster
initial decay can be explained by the larger value of $b$ whereas the slower decay
from the $F_1$=1 to the $F_1$=2 set of hyperfine levels than for \forD~can be rationalized by the 
slightly larger energetic separation and the reduced number of levels of \forH.

In Figure~\ref{fig:14}(b), the density of \fifD~molecules in the trap is shown as a 
function of time. In this case, the decay in the zero field trap is seen to be almost
single exponential, and the 1/e time constant due to the nonadiabatic
transitions is 1.2~s. A more pure single 
exponential behaviour can indeed be expected as all low-field seeking 
hyperfine levels behave very similar in the electric field (see Figure \ref{fig:Stark}(d)). When all voltages 
on the trap (and thus also the electric field offset in the non-zero
field trap) are reduced by a factor of four, the lifetime in the trap with the offset electric
field stays the same (data not shown). In the zero field trap, however, the decay is not 
only faster due to the larger value of $b$ but it is also multi-exponential as the
grouping of the hyperfine levels is now relatively more important.

As evidenced by the experimental results presented in this work, nonadiabatic 
transitions can be a dominant loss channel in electrostatically trapped molecules,
even at temperatures of tens of milli-Kelvins. These trap losses can be effectively 
prevented by using an electrostatic trap with a non-zero electric field at the center, or, 
for instance, by orbiting the molecules off-center in a storage ring \cite{Crompvoets:Nat411:174}. 
The importance of trap losses due to nonadiabatic transitions depends on the 
detailed energy level structure of the molecule under consideration. 
Contrary to the situation in magnetic fields, there is not necessarily a degeneracy of 
low-field and high-field seeking levels for molecules at zero electric field. As a consequence, 
molecules in levels that are exclusively low-field seeking, like \forH~and \forD~molecules 
in the $F_1$=1 hyperfine manifold of the \JK~level (see Figure \ref{fig:Stark}(b,c)), 
can be rather immune to nonadiabatic transitions. For some applications
it might not be desirable to introduce additional offset fields. In that case it might be possible
to select another isotopologue with a slightly different hyperfine
structure to effectively suppress the
losses due to nonadiabatic transitions.  

\begin{acknowledgments}
We thank Henrik Haak for the trap design and for technical support,
and Steven Hoekstra and Andreas Osterwalder for fruitful discussions. M.S. acknowledges a Liebig 
grant from the \textit{Fonds der chemischen Industrie}. 
\end{acknowledgments}

$^*$corresponding author: schnell@fhi-berlin.mpg.de

\bibliography{string,mp}
\bibliographystyle{jk-apsrev}
\end{document}